\documentstyle[12pt,aasms4]{article}

\begin{document}

\title{Stellar iron abundances : non-LTE effects}
\author{F. Th\'evenin, T.P. Idiart}
\affil{Observatoire de la C\^{o}te d'Azur \\
B.P. 4229,F-06304 Nice Cedex 4, France}
\affil{Universidade de S\~ao Paulo, IAG, Depto. de Astronomia
\\ C.P. 3386, S\~ao Paulo 01060-970, Brazil}
\authoremail{thevenin@obs-nice.fr, thais@iagusp.usp.br}

\begin{abstract}

We report new statistical equilibrium calculations for Fe I and Fe II
in the atmosphere of Late-Type stars. We used 
atomic models for Fe I and Fe II having respectively 256 and 190 levels, as
well as  2117 and 3443 radiative transitions.
Photoionization cross-sections are from the Iron 
Project. These atomic models were used to investigate NLTE
(non local themodynamic equilibrium) effects in
iron abundances of Late-Type stars with different atmospheric parameters. 
We found that most Fe I lines in metal-poor stars 
are formed in conditions far from LTE (local thermodynamic equilibrium). 
We derived metallicity corrections of about 0.3 dex with respect to LTE values, 
for the case of stars with $\rm [Fe/H]\sim-3.0$. Fe II is found not to be 
affected by significant NLTE effects.
The main NLTE effect invoked in the case of Fe I is 
overionization by ultraviolet radiation, thus classical 
ionization equilibrium is far to be satisfied. An important consequence
 is that surface gravities derived by LTE analysis are in error and 
should be corrected before final abundances corrections.
 This apparently solves the observed discrepancy between spectroscopic surface 
gravities 
derived by LTE analyses and those derived from Hipparcos parallaxes.
A table of NLTE 
[Fe/H] and $\rm log\, g$ values for a sample of metal-poor late-type stars is 
given. 
\end{abstract}


\section{Introduction}

Iron is a basic stone for the study of the chemical evolution of stellar
systems. Relations
between elemental abundance ratios [X/Fe] versus [Fe/H] 
\footnote{X represents elements heavier than He}  
of giant and dwarf stars are generally used as tracers of chemical evolution 
of galaxies. The reason of this choice is that Fe lines are often
 quite numerous and easy to detect, even in very metal-poor stars. 
Hence, a good and precise determination of Fe abundances is of
 fundamental importance. 
Our understanding of Fe abundance in stars is mainly based on local
 thermodynamic equilibrium (LTE) analyses, for which many  weak Fe I 
and Fe II lines are used. Most of the works devoted to spectral abundance
 analysis assume that the majority of these 
weak lines are not affected by important LTE deviations, in particular
in the case of solar type dwarf stars. However, empirical studies of the 
ionization equilibrium of Fe I show some evidences that important NLTE 
(non local themodynamic equilibrium) effects
are suspected to be present in the photosphere of very metal-poor stars 
(Zhao \& Magain 1990, Fuhrmann 1998, Feltzing \& Gustafsson 1998).
Calculations performed by Takeda (1991), which refers
largely to other excellent computational works in the literature, imply in 
marginal NLTE effects on the ionization equilibrium
for the giant star Arcturus and more important ones when removing the 
UV line opacities. NLTE effects were also investigated by Gigas (1986) in the 
hot star Vega, finding a correction of 0.3 dex for the iron abundance with
respect to LTE value. Rutten (1988) and Steenbock \& Holweger (1984) have also
enlighten the problem, determining the most important NLTE mechanisms 
working in stellar atmospheres of Late-Type stars.
 If important corrections should have to be applied to LTE abundances, 
this could have a real importance on galactic chemical evolution models.

The basic theoretical problem of determination of stellar abundances through 
high resolution spectral analysis is to solve the equation of radiation 
transfer, giving the variation of the radiative energy 
flow throughout an absorbing and emitting gaseous medium. Firstly, the 
solution requires the construction of a stellar atmosphere model, which gives
 the thermodynamic variables, temperature $\rm T$ and pressure $\rm P$,  as a 
function of the optical 
depth $\tau $. Secondly, we  have to determine the absorption and emission
coefficients $\rm j_{\nu}$ and $\rm k_{\nu}$, which are directly proportional to 
the
transition probabilities and the number of atoms, ions or molecules in a
given quantum state or energy level. Atomic, ionic and molecular populations
 depend on the elemental abundance and on the degree of  excitation, 
ionization 
and dissociation respectively. These last quantities are calculated through the
solution of 
the statistical equilibrium equations for given conditions of abundances, 
gas density, temperature, depending also on atomic and molecular constants. 
In a NLTE problem, elemental populations will depend, not only on 
gas temperature and density (as in LTE case), but also on the radiation field. 
To take this into account, we must analyze each line transition 
by considering the transitions of neighbouring levels and not as a two-level
atom as in the LTE case. As a consequence, we have to calculate simultaneously 
the 
transfer and the statistical equilibrium equations for each considered level.
  
The real difficulty to treat a complete iron atom in the NLTE case, is 
to solve 
the statistical equilibrium equations including about 300 terms and 5000 
multiplets. The diagnostic concerns a lot of strong UV lines, 
a lot of optical lines (as is used in classical detailed analysis), and also
a lot of infrared lines which come from the highest terms of the atomic model,
i.e., those with an excitation potential greater than 3 eV. We present 
in this work a much more 
complex model for Fe I and Fe II, taking into account a larger number 
of levels and transitions than those found previously in the literature. 
Our main goal is to estimate NLTE effects in the determination of 
Fe abundances
using the curve of growth technique, as in a classical LTE analysis. In
addition, this procedure allow us to check the ionization equilibrium and, 
consequently,
to evaluate NLTE effects in the $\rm log\, g$ determination.

In section 2 we present our atomic models and the strategy to correct Fe 
abundances, 
and in addition surface gravities, of NLTE effects. In sections 3 and 4 we 
discuss 
the results for 
the Sun and stars previously studied in other works, respectively. In section 5,
our results for metal-poor stars are given, and in section 6 we present our
conclusions and some suggestions 
for future stellar abundance analyses.

\section{The iron atom model and the strategy}

\subsection{Fe I and Fe II atoms}

Our first step for the analysis of  NLTE effects was the elaboration 
of Fe I and Fe II atomic models. Our goal was to 
construct the best models from the statistical point of view, 
taking into account a complete set of levels and transitions. 
We have not tried to produce synthetic spectra to compare to observed
profiles,
but to give differential abundance corrections using equivalent widths and 
curves of growth, as will be described later in section 3.

For Fe I and Fe II we consider levels with principal quantum numbers n=1,3,5, 7 
and 
n=2,4,6,8 respectively.  
For computing time reasons we restricted 
our Fe I /Fe II models to 256/190 levels, having a potential less than 6.8/8.7 
eV
and a continuum at 7.90/16.1 eV. In both models, fine structure were 
taken into account, resulting in a total of 2117 radiative transitions for Fe I 
and 
3443 for Fe II. We have not considered line transitions in the far infrared 
($\lambda > 5.0 \mu m$). Grotrian diagrams of our Fe I and Fe II models are 
given in Fig. 1. 

The code used to solve the equations of statistical equilibrium and radiative 
transfer is MULTI, described by Carlson (1986), which use the operator 
perturbation technique of Scharmer \& Carlsson (1985). We used version 2.2 
(1995). 
Radiation fields for each stellar atmospheric model are computed using
opacities from Uppsala package, including effects of some
line blanketing. 
We did not use the option of the code which produces NLTE background opacities, 
once the models of atmospheres used here are based on LTE calculations (see 2.2) 
and our results do not have such degree of precision for individual line 
profiles yet (see section 3).
As stated above, we are interested in the derivation of differential abundance 
corrections 
based on a curves of growth analysis. A better treatment of the opacity would 
be its derivation with the same code used to compute the atmospheric model, 
however, 
such treatment would probably not change strongly our results, mainly for
metal-poor stars. 

\noindent Input atomic data for this code are:

\noindent a) Energy levels:
\begin{itemize}
\item excitation potentials

\item statistical weights

\item ionization stages

\end{itemize}

\noindent b) Transitions:
\begin{itemize}
\item oscillator strengths 

\item radiative and collisional damping coefficients

\item photoionization cross-sections

\item excitation and ionization collisional cross-sections. 

\end{itemize}

Excitation potentials of the levels and their statistical weights are given 
by Hirata \& Horaguchi's table (1995).
For radiative transitions we used oscillator strengths given by 
Fuhr, Martin \& Wiese (1988), Hirata \& Horaguchi (1995) 
and Th\'evenin (1989, 1990).
Damping coefficients are used for calculation of line profiles 
(line broadening). The total line damping coefficient is given by
$\rm  \gamma = \gamma_ {rad }+\gamma_{coll}$, where $\rm \gamma_{rad}$ and
$\rm \gamma_{coll}$ are respectively the radiative and collisional damping 
coefficients. Radiative or natural damping width is defined as :
$\rm \gamma_ {rad } = \sum_{l<i} (A_{il}) + \sum_{l<j} (A_{jl})$
where $\rm A_{il}$ are the Einstein coefficients.
The collisional damping coefficient is the sum 
of van der Waals and 
Stark coefficients, which take into account effects due to perturbations with 
neutral 
H and He (van der Waals) and charged particles (Stark). For all lines the 
classical van der Waals damping is evaluated from the classical approximation 
(Uns\"old 1955) and multiplied by an enhancement factor, since van der Waals 
constant cannot reproduce the real lines profiles (see Gurtovenko \& Kostic 
1982, Th\'evenin 1989 or more recently Anstee, O'Mara \& Ross 1997),
which will be discussed in section 3. 
In our case, damping due to Stark effect can be neglected, because the 
electronic 
density in the photosphere of the Late-Type stars investigated here is much 
smaller 
than neutral H density.

Photoionization for all levels of Fe I and Fe II was treated in details by using 
the 
frequency dependence of the cross-sections given by the Iron Project 
(Bautista 1997). At each layer of the star's photosphere models  
ionizing radiation fields are computed, giving an estimate of 
photoionization rates. Note that if the ionization equilibrium
is not achieved, the opacity computed would have to be changed by changing
the iron abundance.
The very detailed cross-sections given by the Iron Project were smoothed to 
decrease the number of points in tables cross-section vs. frequency for each 
level, 
in order to reduce the computing time. It was easy to check on some levels that 
this 
smoothing of the detailed cross-sections had no consequences on the final 
results 
of the radiative transfer computations. But, for some of the levels, when strong 
resonances in the photoionization cross-sections near the threshold exist, 
details were considered. 

Collisional ionization rates with electrons are derived from the approximate 
formulae 
given by Mihalas (1978). For the evaluation of collisional rates of permitted 
transition
lines we used van Regemorter's formulae (1962). For some forbidden lines 
considered,
collisional rates were derived from the formulae of Auer \& 
Mihalas (1973) with $\Omega$=0.1, not equal to 1.0 as proposed by Takeda (3.1.2, 
1991)(see Cayrel et al. 1996). However, these forbidden lines seem to play a 
negligible role in our Fe models. 
Uncertainties in the collisional processes with neutral H and He are probably
the main source of errors in our models. The lack of accurate cross-sections 
for collisions with hydrogen atoms is 
largely discussed in the recent literature. The use of the Drawin's theory 
(1968, 1969),
as proposed by Steenbock \& Holweger (1984), was severely criticized by
Severino, Caccin \& Gomez (1993), estimating that Drawin's theory gives 
cross-section values 
larger by a factor of $10^3$ for NaD lines. A clear sum-up of the situation is 
given by Holweger (1996).
In the absence of a more precise 
theory we used this approach as already done in many other works. 
But, we emphasize that we found globally no important 
consequences on the population distribution in atomic levels and therefore
on the equivalent width of lines transitions of our Fe I and Fe II models,
when using or not hydrogen collisions. 

We notice that changing Bautista's photoionization cross-section values by 
a factor 2 induces changes in the resulting populations,
giving errors on [FeI/FeII] less than 0.02 dex. This
means that our strategy developed around the technique of the curve of growth
(see 2.3) has a low dependence on such atomic parameters. Of course this 
is not true if profiles of Fe lines are used to perform 
the stellar chemical abundance analyses.

\subsection{The atmospheric models}

In order to investigate NLTE effects on the ionization balance of iron
with different stellar atmospheric 
parameters, atmosphere models for different known stars were generated 
using the Bell et al. (1976) grid. This grid was used to be self-consistent 
with Th\'evenin's catalogue (1998), 
which reanalyzes LTE abundances of 35 chemical elements for 1107 stars. 
These models reproduce well the atmosphere of Pop II F to K dwarfs and giants, 
for which LTE atmospheric parameters were taken from Th\'evenin's catalogue.  
Microturbulent velocities were taken independent of the optical depth. 
No macroturbulent velocity was used in our computations.

We have check that Bell et al.'s models (1976) and Kurucz's models (1993) 
for a dwarf metal-poor star give no significant differences on results of
[FeI/FeII].

The solar iron abundance adopted in this work is $\rm A_{Fe}=7.46$ 
\footnote{$\rm A_{Fe}=
log(N_{Fe}/N_{H})+12.0$}
(Holweger 1979) as also used by Th\'evenin (1998). 
The reason of this choice is clearly explained in Th\'evenin (1989, see
discussion \& Table III). Other determinations have lead to similar solar 
abundances 
(see for example Holweger, Heise \& Kock 1990 or Holweger et al. 1991).

\subsection{The strategy to estimate NLTE effects}

To compute NLTE effects on the derived LTE surface 
gravities and [Fe/H] values, we constructed NLTE curves of growth
for both Fe I and Fe II, using the equivalent widths $\rm W$ calculated with 
MULTI for a given stellar atmospheric model and for all transitions considered, 
as a simulation of observed equivalent widths . The abscissae  
of the curves of growth are computed under the condition of
LTE as classically done by detailed analysis, for each of the 2117 Fe I and 
3445 Fe II lines. However, to fit the classical LTE curves of growth for each 
analyzed star, we used only lines ranging between $\lambda$$\lambda$ 2200-10000 
$\rm\AA$ to be consistent with classical detailed analysis procedure. 
Once that atmospheric parameters estimated for a given star are from LTE 
analysis (see Table 2, Th\'evenin 1998), we can immediately 
check the validity of this assumption and correct, if necessary, LTE values of 
[Fe/H] in the same table. But, before to correct the metallicities, 
we have to check the ionization equilibrium to derive the error on log g 
due to NLTE effects. If [FeI/H] were not equal to [FeII/H],
this would mean that the classical LTE detailed analysis is wrong and
would have forced the ionization equilibrium using a wrong surface gravity.
Hence, for all stars, we had to calculate theoretical curves of growth with 
new corrected values of $\rm log\, g$ and then estimate metallicity 
corrections.

\section{Results in the solar case}

For the Sun, we used a model from the grid of Bell et al. (1976) given by 
Gustafsson (1981) and, in addition, tested the Holweger-M\"uller's model (1974) 
to compare the results of both models for profiles of strong lines. We used the 
same solar abundance 7.46 for both models. Defining $\rm W_{NLTE}$ and $\rm 
W_{LTE}$ as
the computed equivalent width in NLTE and
LTE conditions respectively, results of  $\rm W_{NLTE}/W_{LTE}$ ratios, which 
give the importance of NLTE effects on each spectral line,
differ but not considerably from one 
model to other. However, these two models give differences on the profile of 
strong lines like $\lambda$ 4045 \AA \, (see Fig. 2).
The enhancement factor of the 
$\rm \gamma_{H}$ van der Waals constant has to be 2.5 for Holweger-M\"uller's 
model and only 1.3 when one uses the Gustafsson's model, to fit perfectly the 
profiles by Kurucz et al. (1984). The value of 2.5 is classical and universally 
adopted for most of the lines, but it can vary from one line to another. 
This problem has been treated by Anstee, O'Mara \& Ross (1997) 
who reproduce well enhancement factors
ranging from 1.4 to 3.3. Once that for late-type 
stars analysis we used the Bell et al.'s grid of models, we decided to keep the 
value of 
1.3 for all radiative transition lines of our Fe I and Fe II models, since we do 
not try to reproduce perfectly all the line profiles.
This problem was also found for the Ca I triplet lines
(e.g. Cayrel et al. 1996), 
where Gustafsson's model looks less good when using a theoretical 
enhancement factor of 2.44 for the
Ca I line $\lambda$ 6162 \AA, compared to the result obtained with the 
Holweger-M\"uller's model. This of course has an important consequence on 
detailed 
analysis of very metal-poor stars, because these strong lines would have to be 
used
once that they are the only ones measurable on observed spectra. 

For solar photosphere, there were no important NLTE effects (overionization 
around 0.02 dex) found by the position of the curves of growth of Fe I and Fe 
II. Once the precision of curve of growth's fit is around
0.04 dex, we have adopted 0.0 dex NLTE effect for  
Gustafsson's solar model.
Consequently, one can say that the ionization equilibrium is perfectly 
reproduced by the LTE surface 
gravity $\rm log\, g_{\odot} = 4.44$. This absence of overionisation is due to
important UV line blocking in solar dwarf stars and
is independent of the solar model used. Fig. 3 shows the solar theoretical 
curves of growth of Fe I and Fe II considering only computed lines ranging 
between $\lambda\lambda$ 2200 - 10000 \AA \,(see section 2.3).
As we can see, curves of growth have a thickness which
is produced by small NLTE effects giving an increase or a decrease of $\rm W$,
depending on population of levels from which lines are formed - e.g. depending 
on the excitation potential (see also Holweger 1996). These effects can be 
seen by the analisys of each line profile, whose detailed study is not the 
purpose of this paper, as mentioned in section 2.1. Our interest is
centered in the problem of overionization, which have direct consequences on
the ionization equilibrium (not in NLTE effects in line profiles)
and therefore on curve of growth analysis.

In 1949, Carter was the first to point out that the solar curve of growth in its 
damping part is divided into two main branches. 
Different conclusions were proposed to interpret it: the odd-even effect, 
the dependence of the damping constant on the multiplet of the lines 
(Carter 1949, Cayrel de Strobel 1966, Pagel 1965), but no comprehensible 
solutions were demonstrated (Foy 1972). Rutten \& Zwaan (1983)
had suspected NLTE effects.
Usually neglected, this effect had no disastrous 
consequences on curve of growth 
analysis, because the lines used to determine stellar abundances were not  
strong enough to lie in the damping part where this effect exists.
Fig. 3 shows clearly double branches in the damping part for Fe I and Fe II.
We remarked that the lower branch of damping curve of, for example, Fe I 
refers to lines originated from the ground levels (5D). The upper part is 
populated by lines originated from the (5F), which can be
considered as resonance levels, because there are no permitted 
transitions with the ground level (5D). 
It seems that these 5D levels are pumped toward 
upper levels with more efficiency than the next levels (5F). 
UV lines clearly pump these levels as shown
on Fig. 4, where
$\rm b_{i}=n_{i}^{NLTE}/n^{LTE}_{i}$ is the classical coefficient used to 
study NLTE effect. In these two levels, $\rm b_i$
have respectively values close to 0.13 and 0.24. Damping lines vary as 
the square of
abundance, which means that for two lines having the same abscissae (log X)
the ratio of their $\rm W_{NLTE}$ is varying as the square of 
$\rm b_{i}/b_{j}$, i.e. 0.14 as can be checked on Fig. 3. This amplitude of the
double branching of the Carter's effect is of the same order of magnitude
as found on empirical solar iron curve of growth. We drawn attention to
spectroscopists using lines originated from these Fe I resonance levels to
be careful before deriving stellar iron abundances for very
metal-poor stars, for which these lines are the only one easily measurable.
Fe II lines present the same effects of splitting, but NLTE effects 
(see Fig. 4) are globally
less pronounced, because they are formed in deeper parts of the
photosphere. However, Anstee, O'Mara \& Ross (1997) proposed a new approach of 
the 
theory of damping constant which could help to understand 
Carter's effect. Probably more detailed and precise computations are needed
to determine the contribution of both results in Carter's effect.

\section{Comparison with other works}

Before to compute numerous new NLTE surface gravities $\rm log\, g$ and [Fe/H] 
for metal-poor stars, we decided to compare
our technique predictions for two stars
previously studied by other authors. One is Arcturus (Takeda 1991) 
and the other is Vega (Gigas 1986).
In the case of the cool giant metal-poor star Arcturus 
we found negligible NETL effects of the same order as computed by Takeda: 
0.03 and -0.02 respectively for Fe I and Fe II.
For the hot star Vega, significant NLTE effects were derived.  Because NLTE
corrections on abundances of Fe I and Fe II were not the same,
ionization equilibrium was not satisfied by the input of LTE surface gravity.
We had to iterate until reach the following values for 
Vega : $\rm log\, g_{Vega} = 4.29$,
$\rm [Fe/H]_{Vega} = -0.33$ ($\rm A_{\odot}=7.46$).
This iron abundance is comparable to that of Gigas (1986), who 
proposed -0.55 dex (using a solar reference value of 7.67). It should be
noted that most of Fe I lines are formed around $\rm log \tau \sim -1. -2.$, 
on the contrary Fe II
lines are generally formed in deeper parts of atmospheres.

\section{Results on metal-poor subgiant to subdwarf stars and consequences}

We selected in Th\'evenin's catalogue a set of 136 subgiant to subdwarf stars,
with abundances ranging between -4.0 and 0.0 dex. We applied our strategy
described in section 2.3 to each of them. Results are presented in table 1.
As one check of our final results, we compare computed $\rm W_{NLTE}$ for Fe I 
and Fe II lines for the metal 
poor star HD 140283 with those measured by Ryan, Norris \& Bessel (1991).
The correlation is shown on Fig. 7. One can note that Fe I and Fe II 
lines are mixed, meaning that the estimated value $\rm log\, g = 3.74$ is 
correct - same conclusion for
[Fe/H] = -2.21. Its surface gravity was changed from giant to subgiant 
as derived by Hipparcos parallax ($\rm log\, g = 3.79$, Nissen, H{\o}g \& 
Schuster (1997)).We show on Fig. 5 and Fig. 6 the amplitude of the 
overionization in the 
atmosphere of HD 140283 analyzed with the LTE atmospheric parameters.

Nissen et al. (1997) have shown that exists a discrepancy between 
spectroscopic $\rm log\, g$ taken in the litterature 
and those deduced from Hipparcos parallaxes. 
On Fig. 8 are plotted the $\rm log\, g_{NLTE}$ values derived 
by us versus $\rm log\, g$ Hipparcos (Nissen et al. 1997 and Clementini et al. 
1998) for stars in common, showing 
that our results are remarkably close to those from Hipparcos.
Among the four points having the worst correlation (Nissen et al.'s values), 
two stars are suspected to be
double. The adopted error bars are those of Nissen et al. (1997), 
Clementini et al. (1998) and Th\'evenin (1998).

Overionization by UV lines seems to play an 
important role in stellar atmospheres of poor Late-Type stars.
As can be seen on Fig. 6, $\rm b_i$ Fe I coefficients are for most of the levels
far below 1.0, increasing from resonance levels to upper levels
until $\sim 1.2$ - upper levels are concerned by infra-red transitions.
In consequence, the source function $\rm S_{\nu} \approx b_{j}/b_{i} \times 
B_{\nu}$
satisfies the relation $\rm S_{\nu} > B_{\nu}$ for most of UV transitions
(where $\rm B_{\nu}$ is the Planck function). Also for strong resonance lines
the mean intensity satisfies the relation $\rm J_{\nu} > B_{\nu}$ and 
drain lower levels towards upper levels, which are more easily ionized.
Consequently, upper levels are overpopulated,
$\rm J_{\nu}$ decreases far below the Planck function and produces infrared
recombinations. These important mechanisms are well explained in 
Bruls, Rutten \& Shchukina (1992) and we refer to this paper for more details.

The results for our sample of stars are :

- If $\rm T_{eff}$ increases, 
UV radiation field and the pumping of the resonance levels
increase, but the infrared radiation field decreases
too and the overionization do not increase as it would be expected.

- If $\rm T_{eff}$ decreases,
infrared radiation increases and are more efficient to overionize the upper
levels. The effect of $\rm T_{eff}$
variations on the overionization for F to G stars 
is found to be not very pronounced.

- If abundance decreases, UV line blocking decreases rapidly and
the overionization becomes very important as shown on the relation between
metallicity correction factors $\rm \Delta _{[Fe/H]}$ and metallicities 
estimated by LTE approach (Fig. 9).
This reveals that the overionisation increases rapidly when the abundance
decreases from $\approx -0.3$  to -1.5 dex to reach a maximum for metal-poor 
stars having [Fe/H] $\approx$ -3.0. There are no significant overionization
for solar type stars.
                                  
The most important parameter to produce overionization, the main source of
NLTE effects on curves of growth technique, is therefore the variation of
metal abundance in atmospheres of Late-Type stars.  

The precision of $\rm [FeI/H]$ or $\rm [FeII/H]$ determination by using
theoretical curves of growth is estimated to be 0.04 dex, as mencioned in
section 3, meaning that corrections on $\rm log \, g $ reach a precision 
of 0.08 dex.

We computed Fe I and Fe II curves of growth for hot supergiant
star parameters corresponding to stars analysed in the Magellanic Clouds.
These stars have moderate underabundances ($\rm \sim -0.6$ to $-0.2$) dex. 
We found a [FeI/FeII] balance between 0.05 and 0.01 dex for SMC stars (see 
stellar
parameters in Th\'evenin 1998).
Holweger (1990) mentioned 
that iron NLTE effects computed using the LTE atmospheric parameters of the 
poor giant star $CD-38^{o}245$ lead to an error in the
ionization equilibrium balance of +0.3 dex, value very similar to those
derived by us for dwarfs stars.
These results are not surprising as mentioned by Feltzing $\&$ Gustafsson 
(1998), once the UV opacity increases with the increase of the Balmer 
lines when the surface gravity decreases. This only means that the idea 
that NLTE deviations must necessarily increase with the decrease of surface 
gravities (hence increase in collisions) is not entirely correct. From the 
point of view of Boltzmann's law this is true, but if the UV flux
is blocked by increasing of Balmer lines, deviation from
the point of view of Saha's law do not increase with $\rm log g$ decreasing.

\section{Conclusion}

We presented a study of departures from LTE for Fe I and Fe II, mostly for 
metal-poor stars. 
The mechanism of these departure is clearly identified as overionisation, 
responsible for important corrections on the values of surface gravities. 
A good correlation between our derived surface gravities and those deduced
from Hipparcos parallaxes is a proof of the validity of our results in 
table 1, having important consequences on distances in the Galaxy.
These corrections on the surface gravity could have also important incidences
on the abundances of elements like Be (Gilmore, Edvardsson \& Nissen 1991).
NLTE abundance 
corrections are found to be less than 0.35 dex, but not negligible. Therefore,
stellar abundance ratios could have to be revisited after having estimated 
possible 
NLTE effect on other elements like Ca, Mg, Al, O and Be. We are preparing papers 
on these 
subjects.
We recommend to stellar spectroscopists working on metal-poor stars to use NLTE 
computations or the table 1 before to publish their LTE results, or to use 
surface 
gravities derived by Hipparcos parallaxes combined with a LTE analysis of 
Fe II lines which do not suffer important NLTE effects.

\acknowledgements

We wish to thank M. Bautista for having provided his photoionization
cross-sections prior to publication and R. Cayrel, J.A. de Freitas 
Pacheco, R. Gratton,
F. Paletou, J. Tully and Cl. Van t' veer for fruitful discussions. 
This work was facilitated by the use of the
SIMBAD database, operated by the Centre de Donn\'ees astronomiques de 
Strasbourg (CDS - France) and has been performed using the computing
facilities provided by the program {\it Simulations Interactives et 
Visualisation en Astronomie et M\'ecanique (SIVAM)} at the computer
center of the Observatoire de la  C\^ote d'Azur.

T.I. acknowledges the brazilian agencies CNPq for the post-doctoral fellowship
201249/95-2 at the Observatoire de la C\^ote d'Azur during the years 1996-97,
and FAPESP for the present post-doc
grant 97/13083-7 at IAG. F.T. thanks FAPESP for the grant 98/10869-2 at IAG.

\newpage
\figcaption{Grotrian diagrams for Fe I and Fe II atoms. Each 
configuration term is denoted by an alphabetic notation, representing 
the degeneracy of the state. Two Fe II levels do not have configuration 
terms identified, then they were labeled 62065 and 65363 
(which refers to line transitions $\lambda\lambda$ $6206.5$ and 
$6536.3$ \AA) as Hirata \& Horaguchi's table.}

\figcaption{Comparison between two calculated Fe I NLTE line profiles and
observed spectra for the sun. The theoretical profiles were calculated using
the atmospheric models of Holweger-M\"uller (1974) and Gustafsson (1981), for
different enhancement factors $f_{H}$ of the van der Waals constant. Observed 
spectra
are from Kurucz et al. (1984).}

\figcaption{Theoretical Fe I and Fe II curves of growth for the sun.
$\rm log X= log \, gf + log \Gamma + log(N_{Fe}/N_{H}$), where 
$gf$ is the oscillator strength; $\Gamma$ is a parameter defined for each line
as a function of the atmospheric model (Cayrel \& Jugaku 1963); 
$\rm N_{Fe}/N_{H}$ is the relative abundance.}

\figcaption{Departure coefficients of Fe I and Fe II versus optical
depth in the solar case.
$\rm b_{i}=n^{NLTE}_{i}/n^{LTE}_{i}$, where $\rm n^{NLTE}_{i}$ and $\rm 
n^{LTE}_{i}$
are NLTE and LTE populations respectively for each level.} 

\figcaption{Theoretical Fe I and Fe II curves of growth for the star
HD 140283, computed with LTE atmospheric parameters: $\rm \theta_{eff}$=0.90,
$\rm log\, g=3.20$ and $\rm [Fe/H]$=-2.5. For Fe I is overplotted the classical 
LTE
curve of growth (full line), showing the need of correction by NTLE effects.}

\figcaption{Departure coefficients of Fe I versus optical
depth for HD 140283 ($\rm \theta_{eff}$=0.90,
$\rm log\, g=3.20$ and $\rm [Fe/H]$=-2.5).\, $\rm 
b_{i}=n^{NLTE}_{i}/n^{LTE}_{i}$, 
where $\rm n^{NLTE}_{i}$ and $\rm n^{LTE}_{i}$
are NLTE and LTE populations respectively for each level.}

\figcaption{Computed equivalent widths for Fe I and Fe II lines versus observed 
ones by
Ryan et al. (1991) for Fe I and Fe II for HD 140283. Atmospheric parameters 
for calculation of equivalent widths are corrected of NLTE effects.}

\figcaption{Comparison between our derived log g and those using Hipparcos
distances from Nissen et al. (1997) and Clementini et al. (1998). $\rm log \, 
g_{NLTE}$
error bars are estimated from classical detailed analysis for dwarf stars 
$\sim 0.30$ dex (Th\'evenin 1998).}

\figcaption{Final estimated NLTE abundance corrections $\rm \Delta _{[Fe/H]}$ 
versus
LTE atmospheric parameters from Th\'evenin (1998).}

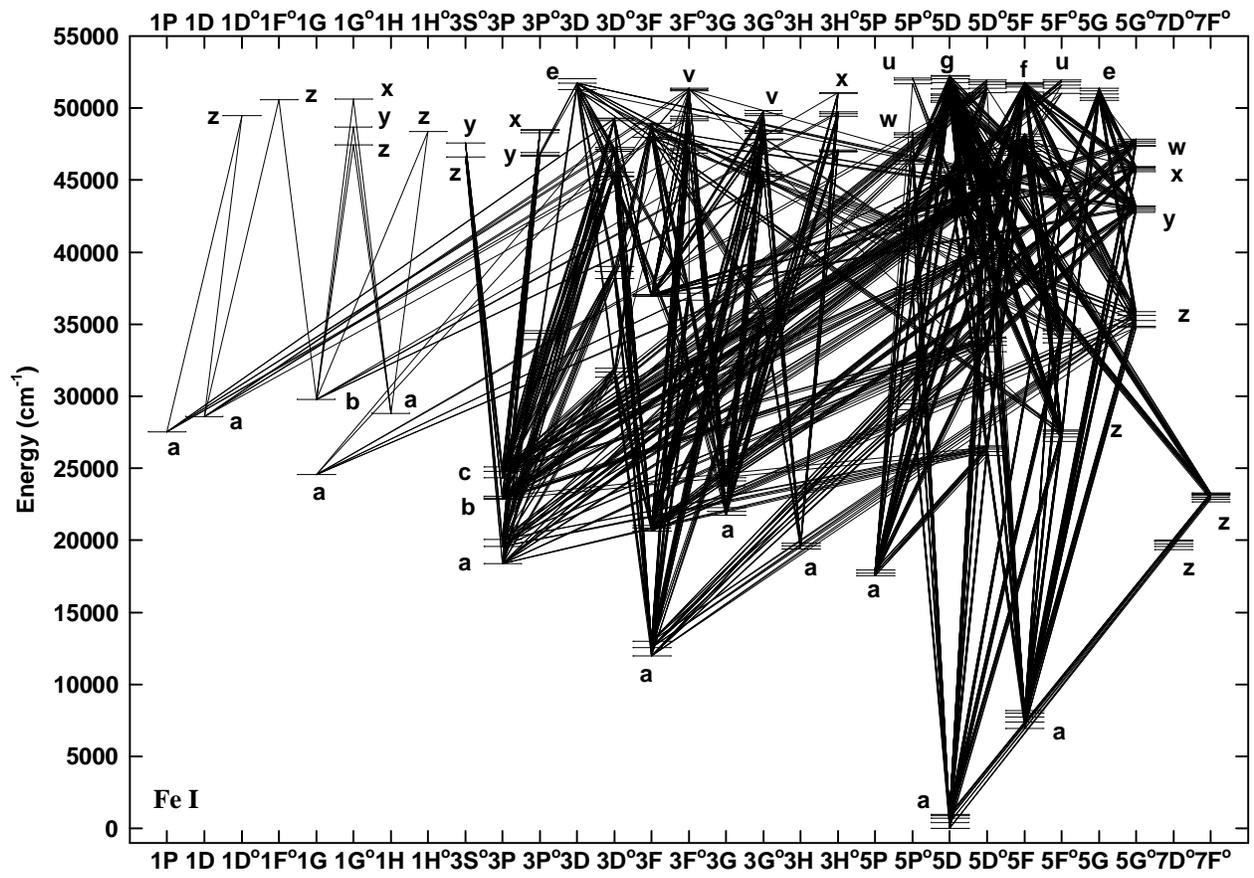
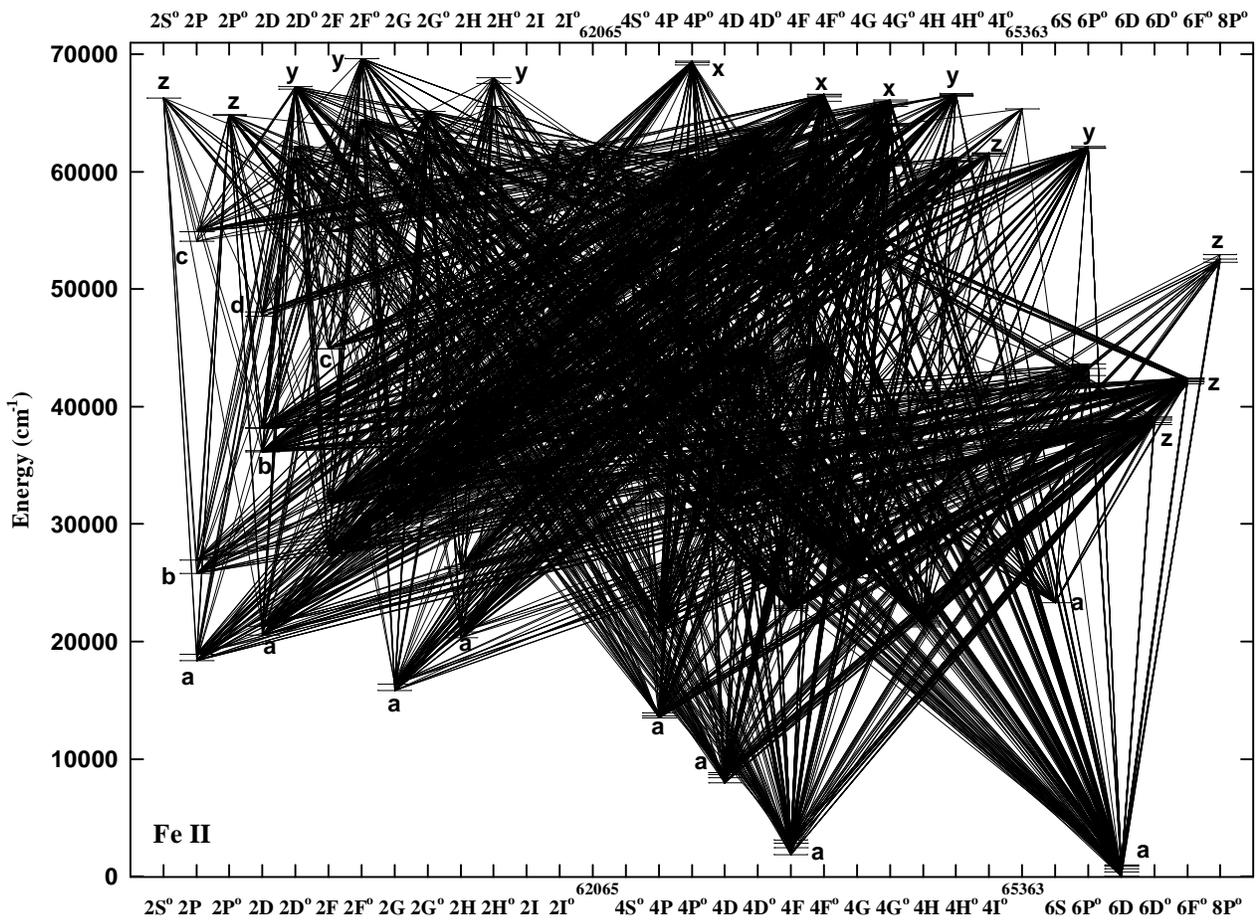
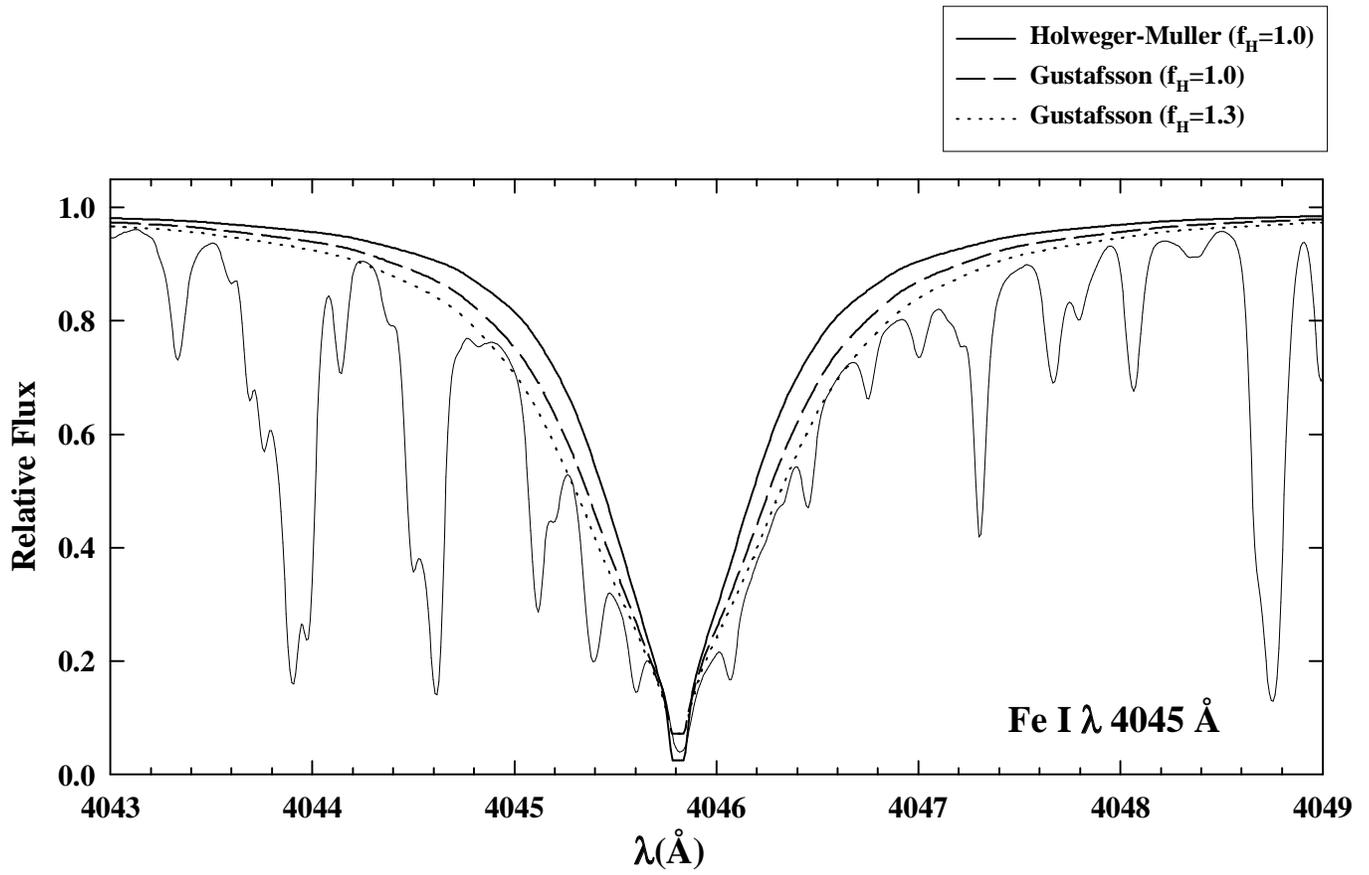
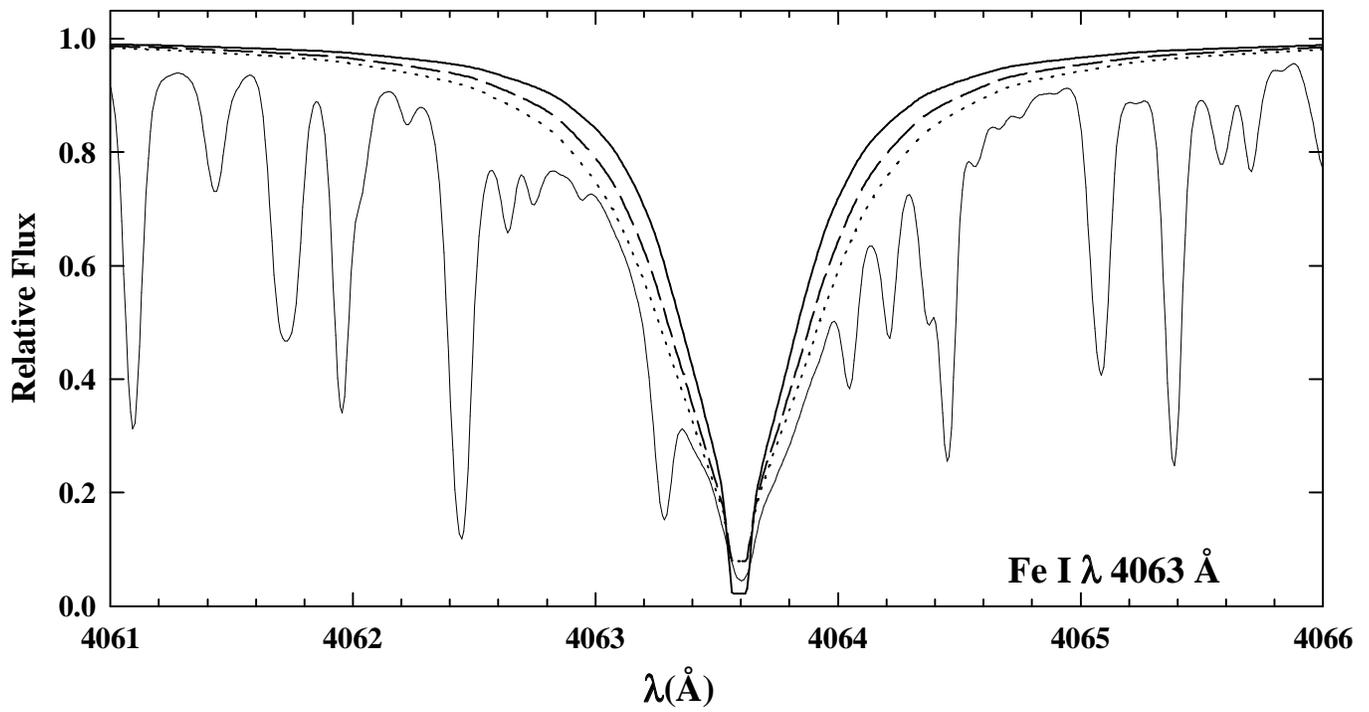
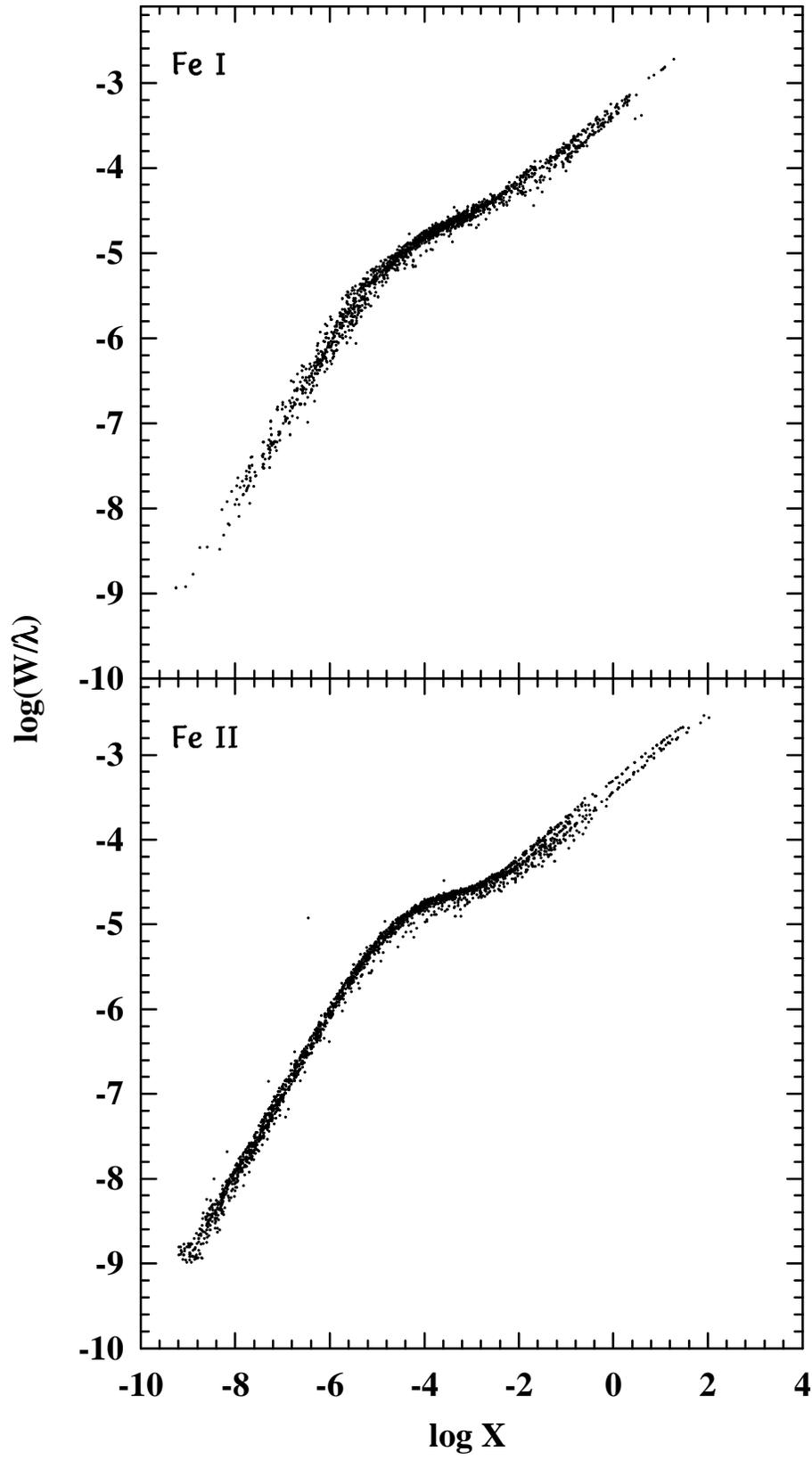
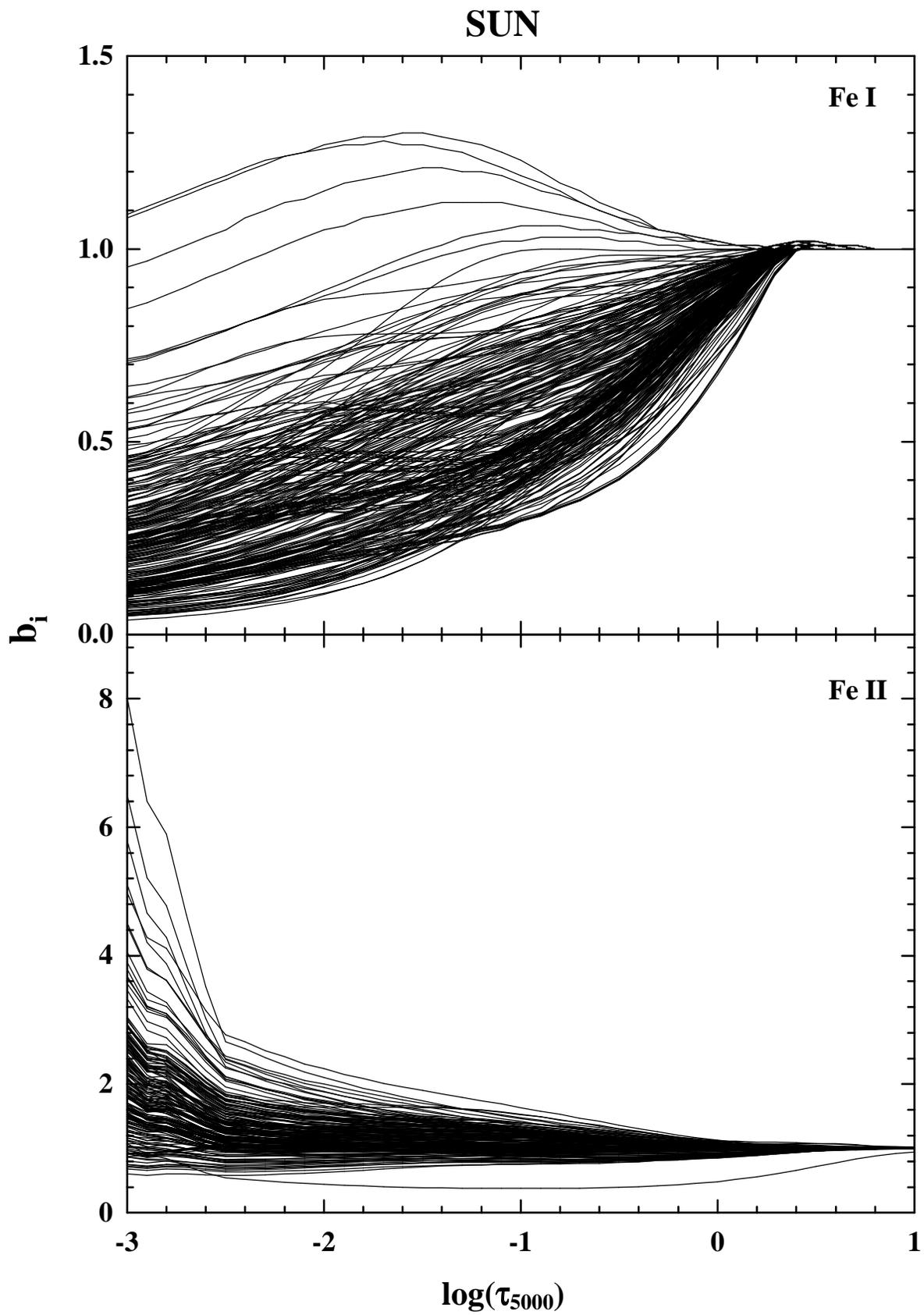
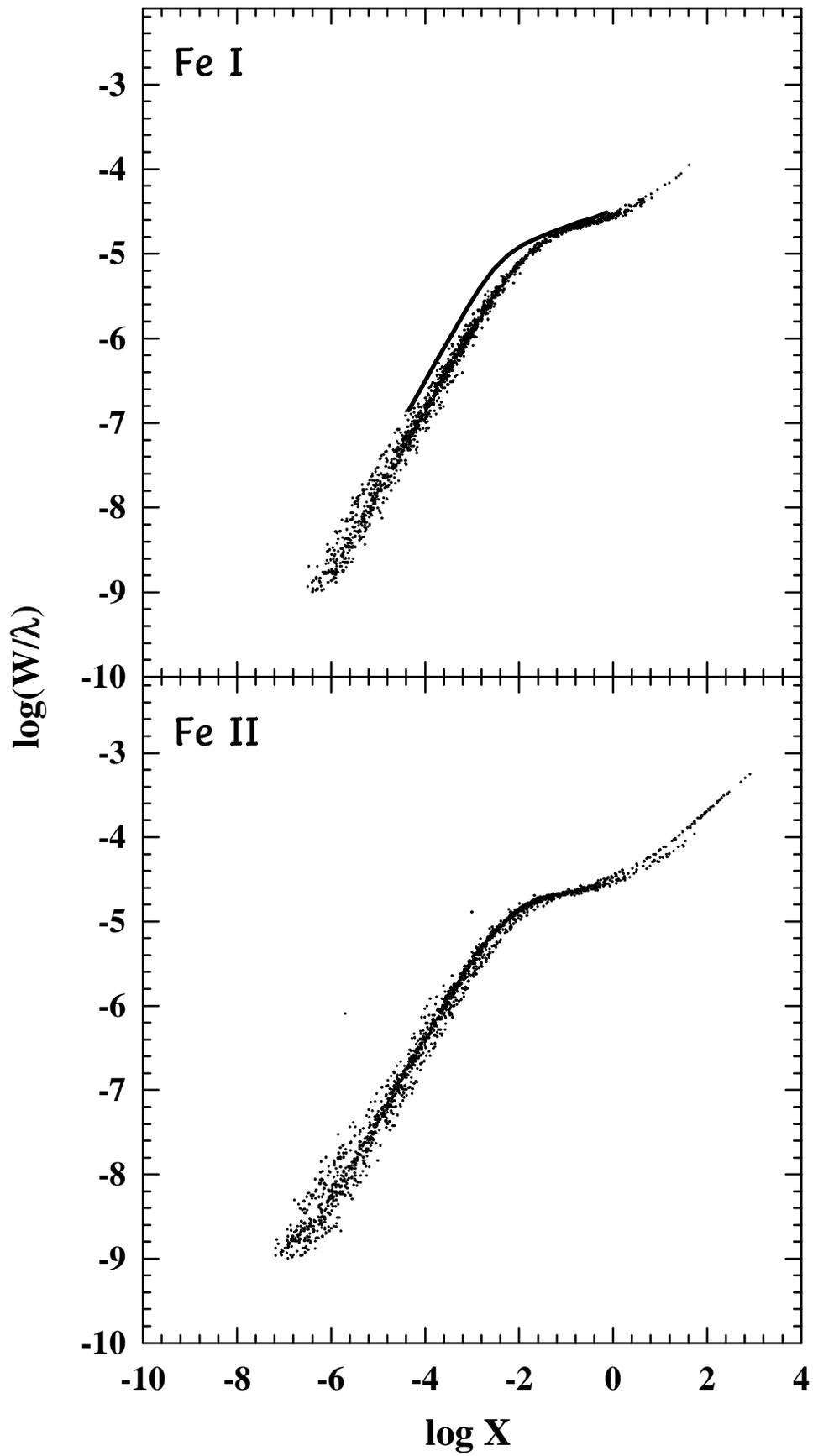
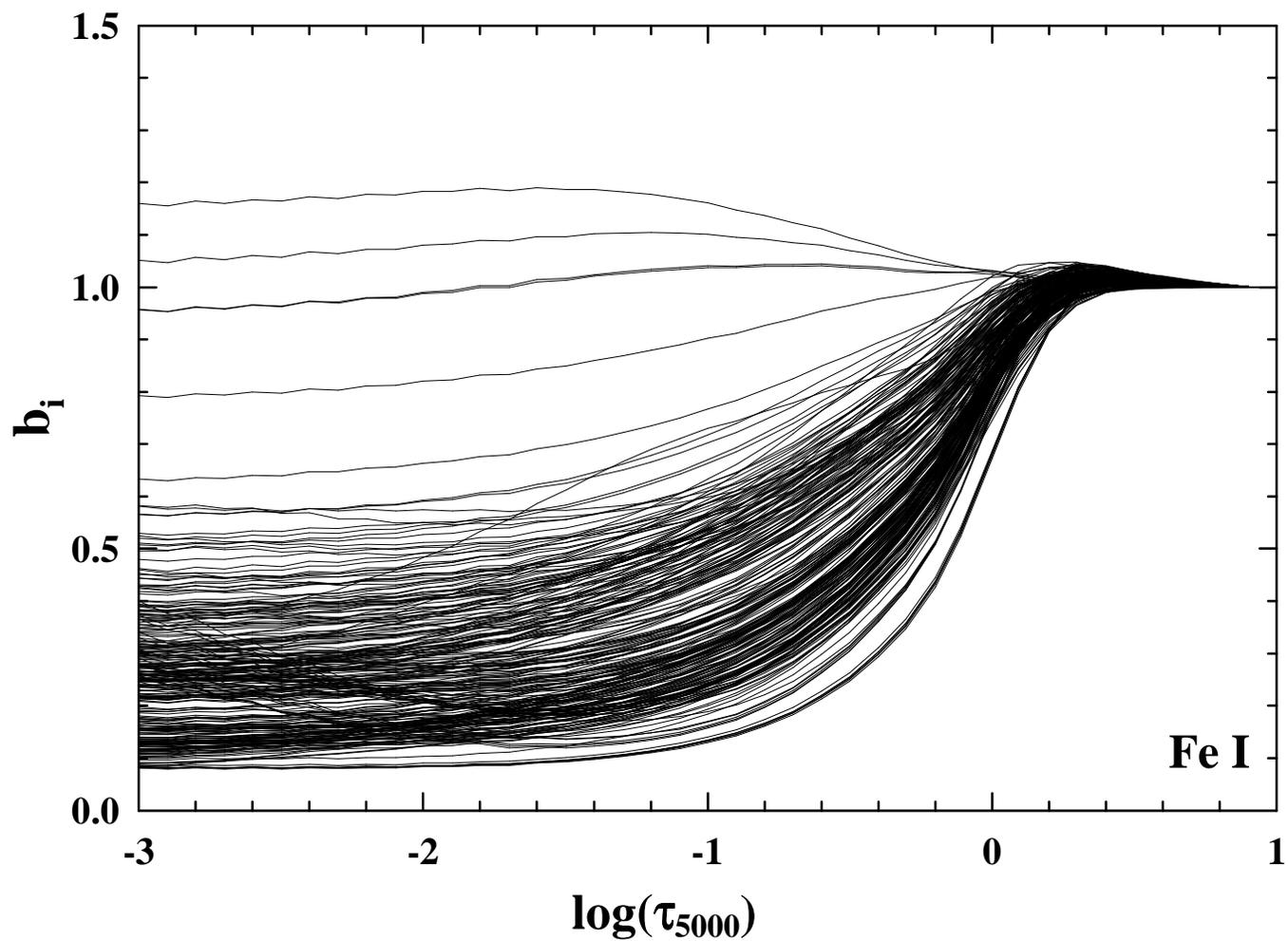
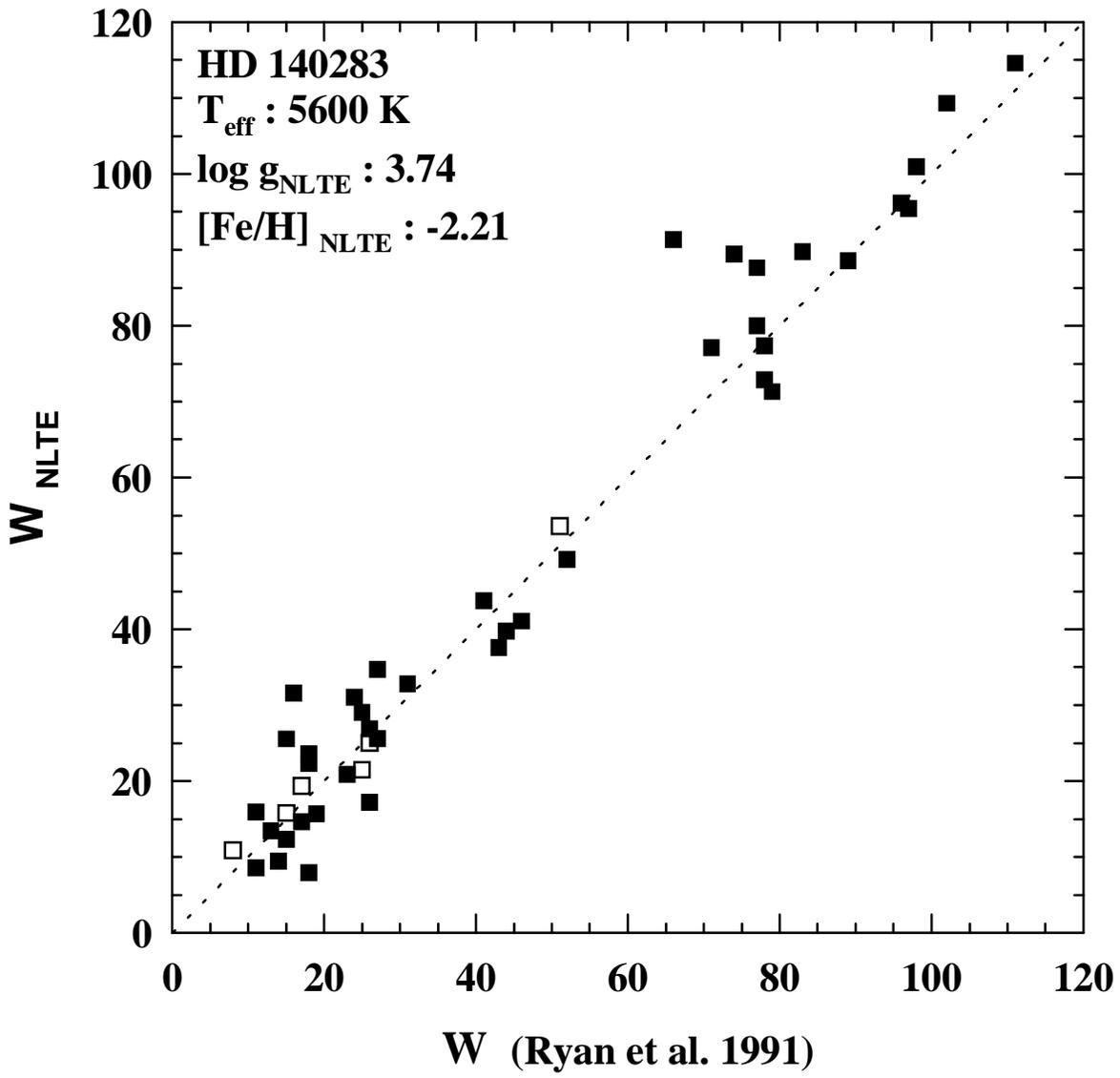
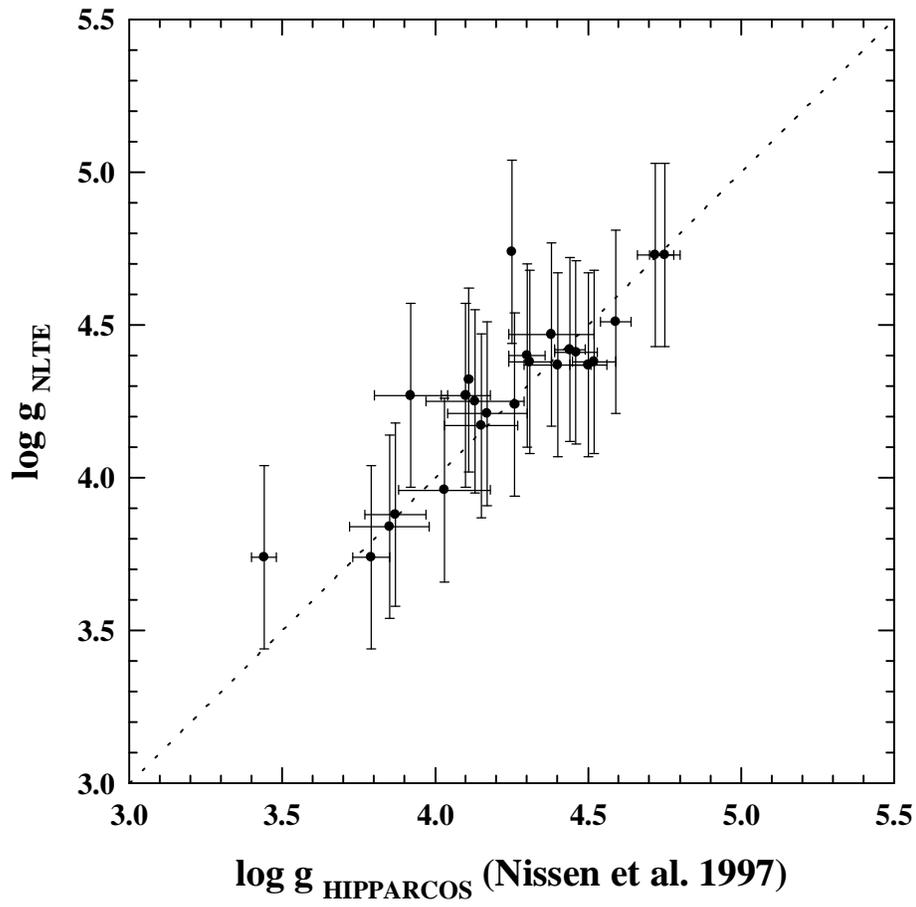
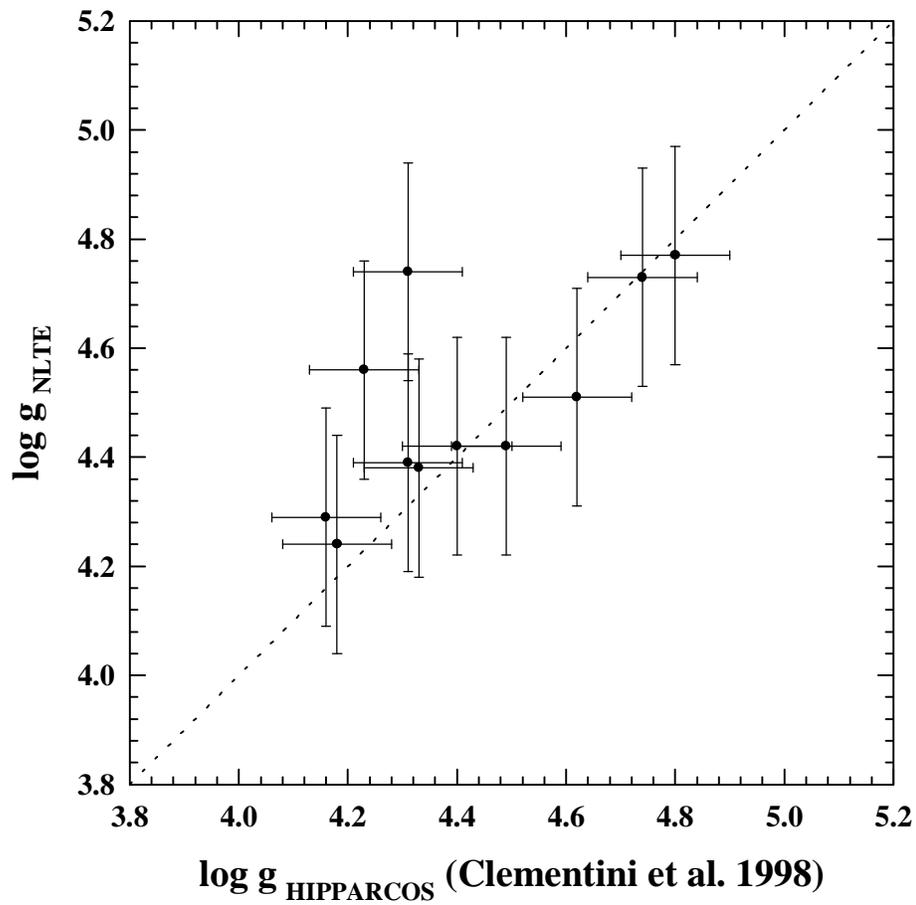
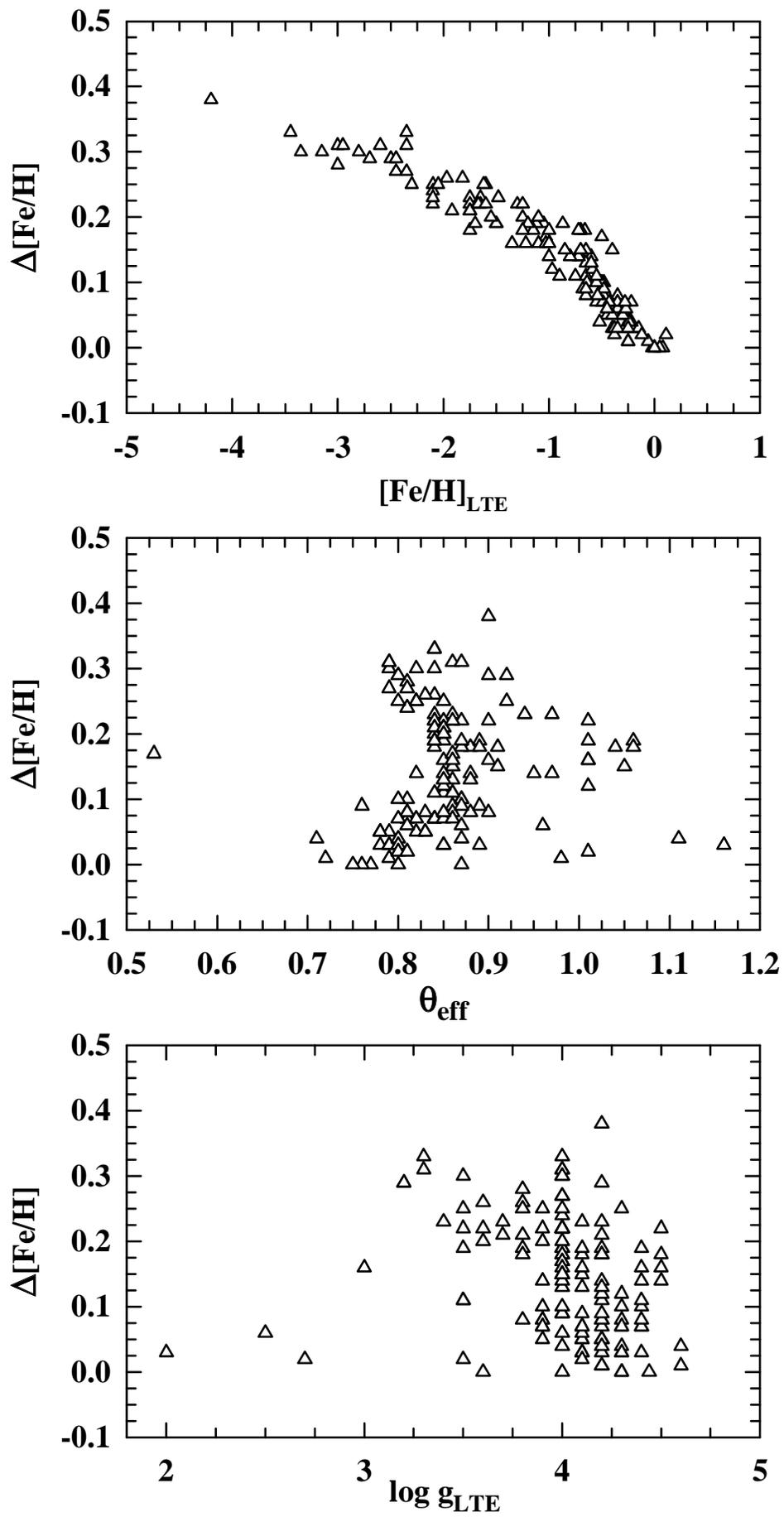
\end{document}